\documentclass[11pt]{revtex4}
\usepackage{graphicx}
\usepackage{amssymb}
\usepackage{amsmath}
\expandafter\let\csname equation*\endcsname\relax
\expandafter\let\csname endequation*\endcsname\relax\usepackage{amsmath}
\usepackage{bm}
\usepackage{multirow}
\usepackage{color}

\begin{document}
\title[Coherent regulation]{Coherent regulation in Yeast's cell-cycle network}
\author{Ne\c{s}e Aral$^1$ and Alkan Kabak\c c\i o\u glu$^1$}
\address{$^1$Department of Physics, Ko\c c University, Rumelifeneri Yolu Sar\i yer 34450, Istanbul, Turkey}

\begin{abstract}
We define a measure of coherent activity for gene regulatory networks,
a property that reflects the unity of purpose between the regulatory
agents with a common target. We propose that such harmonious
regulatory action is desirable under a demand for energy efficiency
and may be selected for under evolutionary pressures. We consider two
recent models of the cell-cycle regulatory network of the yeast, {\it
  Saccharomyces cerevisiae} as a case study and calculate their degree
of coherence. A comparison with random networks of similar size and
composition reveals that the yeast's cell-cycle regulation is wired to
yield and exceptionally high level of coherent regulatory activity. We
also investigate the mean degree of coherence as a function of the
network size, connectivity and the fraction of repressory/activatory
interactions.
\end{abstract}


\maketitle

\section{Introduction}
\label{introduction}
Despite vast amount of data on genetic regulatory interactions in a
growing number of organisms~\cite{yang2014ytrp,
  teixeira2014yeastract,guldener2005cygd, munch2003prodoric,
  drysdale2005flybase, chatr2013biogrid}, the interplay between
structure and function in these systems is still unclear and an active
field of research~\cite{klemm2005topology, mangan2003structure,
  thieffry2007dynamical}. Much effort has been devoted to local
structures and their role in promoting multistationarity, stability
and robustness in a background of fluctuating environmental
conditions~\cite{tyson2010functional,alon2006introduction,
  sevim2008chaotic, cinquin2002roles}. Other studies argue that
regulatory dynamics is ``consistent'' in the sense that, a
transcription factor's (TF) influence on a target gene
(activation/inhibition) is not context
dependent~\cite{sansom2008countering}.  Similarly, architectural
features allowing better controllability have been of recent
interest~\cite{yuan2013exact, yan2012controlling,
  wuchty2014controllability}. Such guiding principles provide valuable
insight and a bird's-eye perspective on networks of gene regulatory
interactions, where a qualitative understanding of the underlying
generative mechanisms is hindered by the prohibitive complexity of the
cell and the shear amount of experimental data.

We here introduce the concept of ``coherent regulation'' as a similar
guideline for the evolutionary design of regulatory networks. Below,
we give a mathematical definition for the degree of coherence, as a
measure of the extent to which a given regulatory network structure
and the equations of motion for the regulation dynamics are optimized
towards minimum waste of material/energy resources available to the
cell. We then calculate it for the yeast cell-cycle network, a
well-studied subnetwork responsible for cell-division in the yeast
{\em Saccharomyces cerevisiae}~\cite{li2004yeast}, by means of two
previously proposed models in the literature. Further analyses on
random graphs yield a reference baseline as a function of the
network's structural parameters, by means of which we show that the
yeast's cell-cycle network is, in fact, optimally organized to
maximize coherent regulatory action.

\section{Model}
\label{model}
\subsection{Coherent regulation}

Regulation of gene activity requires production of regulatory proteins
which demands energy and raw material. The need for the economical use
of these resources has been a constant and dominant evolutionary
pressure in much of the last 4 billion years and virtually for all
species~\cite{zhang2009metabolic, niven2008energy,
  laughlin1998metabolic}. Therefore, it is natural to expect
regulatory networks to have been optimized towards their optimal use.

A full-fetched analysis of such optimization would require a much
deeper understanding than we currently have of a cell's functions on
the protein level. Here, we will discuss network efficiency only in
terms of the choice of interactions that determine the time dependence
of protein expression, and assert that a prerequisite for optimality
is to avoid simultaneous production of regulatory elements that
perform opposing tasks.
In particular, it is desirable that in a steady state, two
TFs where one is an activator and the other is a
repressor of the same gene are not co-expressed. Exceptions are the
relatively brief transition periods (induced by internal or
environmental conditions) between steady states or between the states
in cyclic attractors (such as circadian
cycles~\cite{eelderink2010circadian, akman2012digital}), where changes
in the expression levels are due to role switches between such
antagonistic regulators.

\begin{figure}[b]
	\begin{center}
		\begin{tabular}{|c|c|c|}
			\hline
			&&\\
			\includegraphics[width=2.8cm]{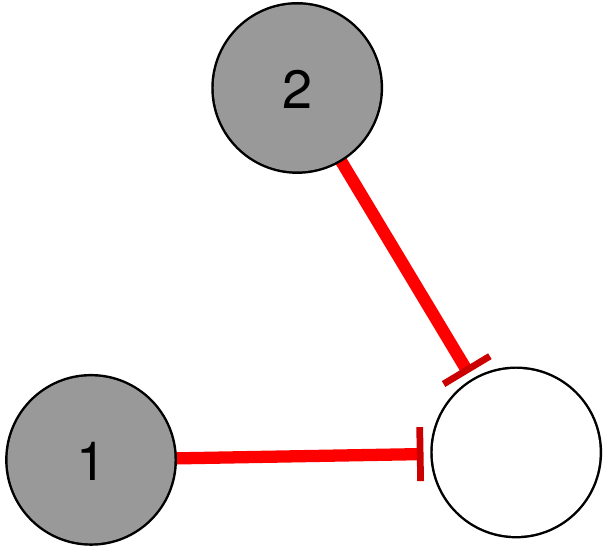} & \includegraphics[width=3cm]{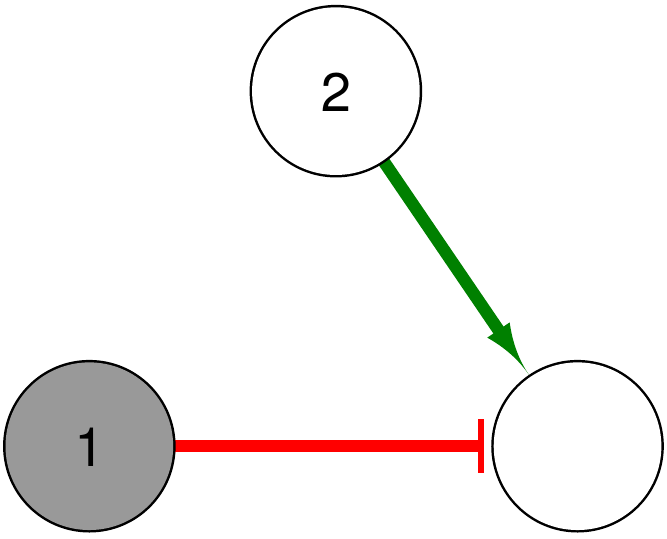} &\includegraphics[width=3cm]{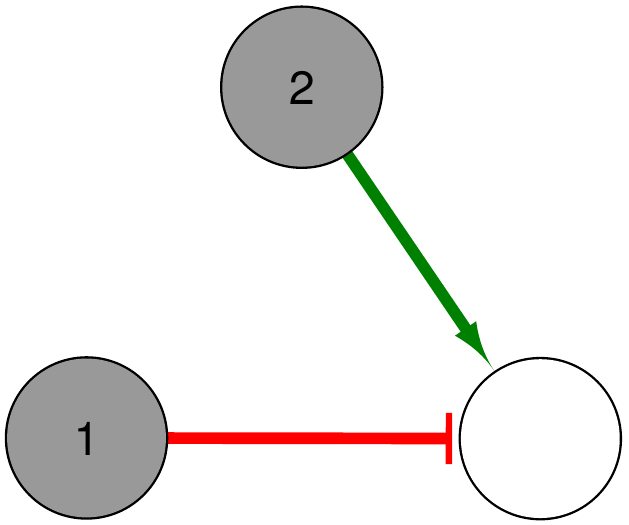}\\
			
			(a)&(b)&(c)\\
			\hline
		\end{tabular}
	\end{center}
	\caption{Coherent vs incoherent regulation for a gene with two
          regulating partners. Active nodes are shown gray. The target
          gene is (a) coherently regulated when the two active
          regulators act unanimously; (b) also coherently regulated
          when the two antagonistic regulators are not active at the
          same time; (c) incoherently regulated when antogonistic
          regulators are simultaneously expressed.}
	\label{coherence}
\end{figure}

We coin the concordance of simultaneously active regulatory elements
sharing a common target as ``coherent regulation''. More precisely, we
say that a gene is coherently regulated if its expression state is
determined unanimously by its active regulatory partners, as described
in Fig.~\ref{coherence}. Note that, coherence defined as above is
different from ``cooperativity'' in regulatory networks discussed in
the literature~\cite{stefan2013cooperative}, which refers to increase
(or decrease) of the binding affinity of a TF when other TFs are
already bound to the promoter region of the gene.
It can also be noted that, a coherent network as defined above is expected to be more
robust against fluctuations in the expression rates of the
co-regulating elements. This is because, when the regulatory inputs
received by a gene are compatible, fluctuations in their relative
strengths (e.g., due to time delays) are less significant than when
the regulatory agents act antagonistically, in which case the
fluctuations may result in a sign change in the net regulatory
message received by the target gene.

The degree of coherent regulation is clearly structure dependent,
i.e., some networks are more supportive of such harmonious regulatory
activity than others. Identifying the architectural aspects of these
networks is an interesting problem in itself which shall be addressed
in a future study.
The efficiency acquired through coherence also has a truly dynamical
aspect, since coherent action is reached through the time-dependent
interplay between expression levels of communicating genes.
Starting from the fact that the structure shapes
regulatory dynamics in a phenotype, the question we ask here is
whether evolutionary processes may have provided the means to feed
information from the dynamics back to the structure, promoting
modifications that maximize coherent regulatory action.

\subsection{Quantifying coherence in regulatory networks}
\label{quantifying}

Given a fixed environment, the genes/proteins that participate in a
genetic regulatory network spend most of their time in the attractors
of the regulation dynamics. Let $\tau(\boldsymbol{\sigma})$ be the
appropriate discrete-time-evolution operator $\tau$ that propagates
the expression level $\boldsymbol{\sigma}$ of the involved
TFs from time $t$ to $t+\delta t$, i.e.,
$\boldsymbol{\sigma}(t+\delta t) = \tau[\boldsymbol{\sigma}(t)]$. Then
the attractors are determined by the condition $\boldsymbol{\sigma}^*
= \tau^q[\boldsymbol{\sigma}^*]$. A point attractor corresponds to
$q=1$.  Otherwise, the attractor is a cycle of period $q\delta t$ with
the cycle states
$\{\boldsymbol{\sigma}^*,\tau[\boldsymbol{\sigma}^*],\dots,\tau^{q-1}[\boldsymbol{\sigma}^*]\}$. We
here consider Boolean models where $\boldsymbol{\sigma}$ is a binary
vector with entries $\in \{0,1\}$.

For stochastic systems, this formulation is easily generalized with,
$\tau\to T$, a transition matrix, and
$\boldsymbol{\sigma}\to\boldsymbol{\pi}$, the probability distribution
on the ensemble of states $\{\boldsymbol{\sigma}\}$. While noise is a
relevant determinant of cell activity~\cite{thattai2001intrinsic,
  swain2002intrinsic, paulsson2004summing}, biological systems are
typically robust to fluctuations in gene expression
levels~\cite{ingolia2004topology, kaneko2007evolution}. Therefore,
both stochastic and deterministic models of regulatory dynamics are
frequently used in the literature~\cite{zhang2006stochastic,
  li2004yeast, faure2006dynamical}.

We can now define the ``coefficient of coherence'' for an attractor
$i$ as $\alpha^{(i)}_{c}$, the fraction of genes which are {\em not}
subject to simultaneous {\em and} conflicting regulatory messages from
their regulating partners active in state $i$. Describing
up-(down-) regulation of gene $j$ by the associated TFs $k_j$ as $c_{jk_j}=1$($-1$), we can express $\alpha_c^{i}$ in
terms of (Boolean) expression levels as:
\begin{equation}
\label{eq:f_i}
\alpha_c^{i} = \frac{1}{N}\,\sum_{j=1}^N int\bigg(\frac{\big|\sum_{k_j} c_{jk_j}\boldsymbol{\sigma}^{(i)}_{k_j}\big|}{\sum_{k_j}  |c_{j k_j}|\boldsymbol{\sigma}^{(i)}_{k_j}}\bigg)
\end{equation}
Above, $N$ is the number of genes in the network. $i$ is the attractor
index, hence for a network with $n_a$ attractors $i$ runs from 1 to
$n_a$. $int()$ function returns the integer part of its argument,
which effectively sets the coefficient of coherence to unity if the
currently expressed subset of regulators $\{k_j\}$ of gene $j$ are all
activators or all repressors, and to zero otherwise. For a cyclic
attractor with period $q_i\delta t$, $\alpha_c^i$ is taken as the
arithmetic average over the cycle states $\boldsymbol{\sigma}_k^{(i)}$
($k=1,..,q_i$). The coefficient of coherence for a network is defined
to be the mean of $\alpha_c^{i}$ over all the attractors $i$, i.e.,
\begin{eqnarray}
\label{eq:coherent}
\alpha_{c} &=& \frac{1}{n_a}\sum_{i=1}^{n_a}\alpha^{(i)}_{c}
\end{eqnarray}
where, again, $n_a$ is the number of attractors. Alternatively, one
could consider the average weighted by the basin sizes (the fraction
of uniformly randomly picked initial states that end up in a given
attractor):
\begin{eqnarray}
\label{eq:coherent_basin}
\alpha_{c} &=& \frac{1}{\sum_{i=1}^{n_a} b_i}\sum_{i=1}^{n_a}b_i\alpha^{(i)}_{c},
\end{eqnarray} where $b_i$ is the number of initial states reaching attractor $i$ under the given dynamics. However, such weighted averaging should be employed with caution, because a
randomly constructed expression state will typically never appear
throughout the life cycle of a cell and, therefore, is irrelevent in
the biological sense. We here present our results with both
basin-weighted and uniformly weighted averages, in order to demonstrate that
our conclusions do not critically depend on this choice.

Finally, the definition given above can also be generalized to
continuous models of gene expression by replacing the averages over
states with time averages and employing an appropriate extension of
the above coherence measure to continuous variables. For simple
linearized models of the form
\begin{eqnarray}
\frac{dn_j(t)}{dt} &=&\sum_{k_j} \gamma_{jk_j}n_{k_j}(t)\\
\end{eqnarray}
where $n_j$ typically represent protein/RNA concentrations, a possible
generalization could be to replace equation~(\ref{eq:f_i}) with


\begin{equation}
\label{eq:f_i_cont2}
\alpha_c^{i} = \frac{1}{T_iN}\sum_{j=1}^N \int_0^{T_i} dt
\frac{\big|\sum_{k_j} \gamma_{jk_j}n_{k_j}(t)\big|}{\sum_{k_j} | \gamma_{jk_j}|n_{k_j}(t)}
\end{equation}
where $T_i$ is the period of the $i$-th dynamical attractor in consideration.

\section{Results}
\label{results}
\subsection{Yeast's cell-cycle: a case study}
\label{yeast}
The budding yeast, {\em Saccharomyces cerevisiae}, is a well-studied
single-cell eukaryote~\cite{kaizu2010comprehensive, hodges1999yeast,
  bader2003functional, teixeira2014yeastract}. Its 307 known TFs out
of $\sim 6000$ distinct proteins participate in (currently estimated)
$\sim 2\times 10^5$ regulatory
associations~\cite{teixeira2014yeastract}. Yet, these are still not
sufficiently well characterized to tell whether an arbitrarily chosen
TF up- or down-regulates a target gene in a given context. Therefore,
the structural optimality -in the above sense- of the yeast's
regulatory network {\em as a whole} is not possible to assess with the
currently available information.

\begin{figure}
\begin{center}
\begin{tabular}{|c|c|}
	\hline
	&\\
	\includegraphics[width=7cm]{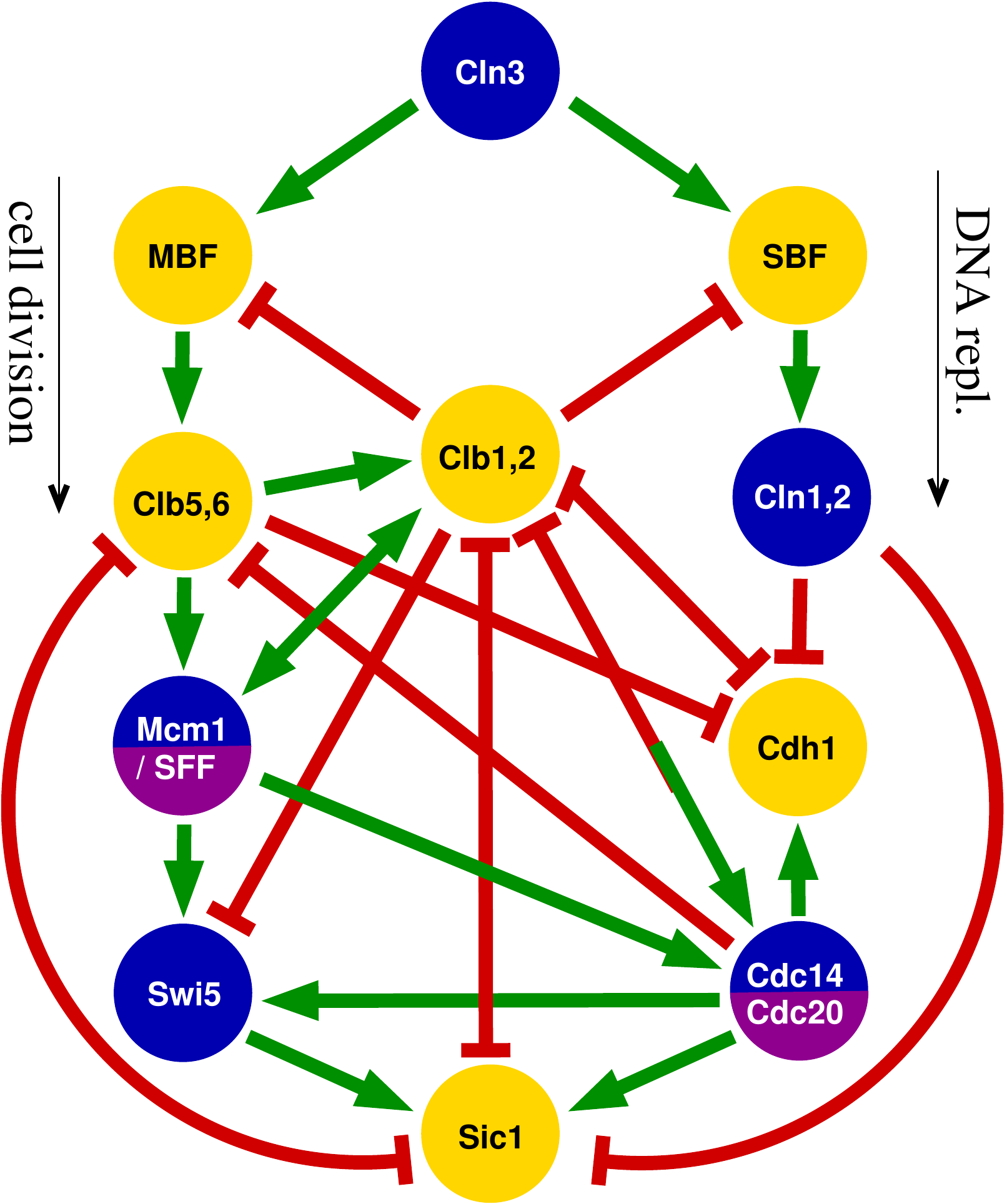}&\includegraphics[width=7cm]{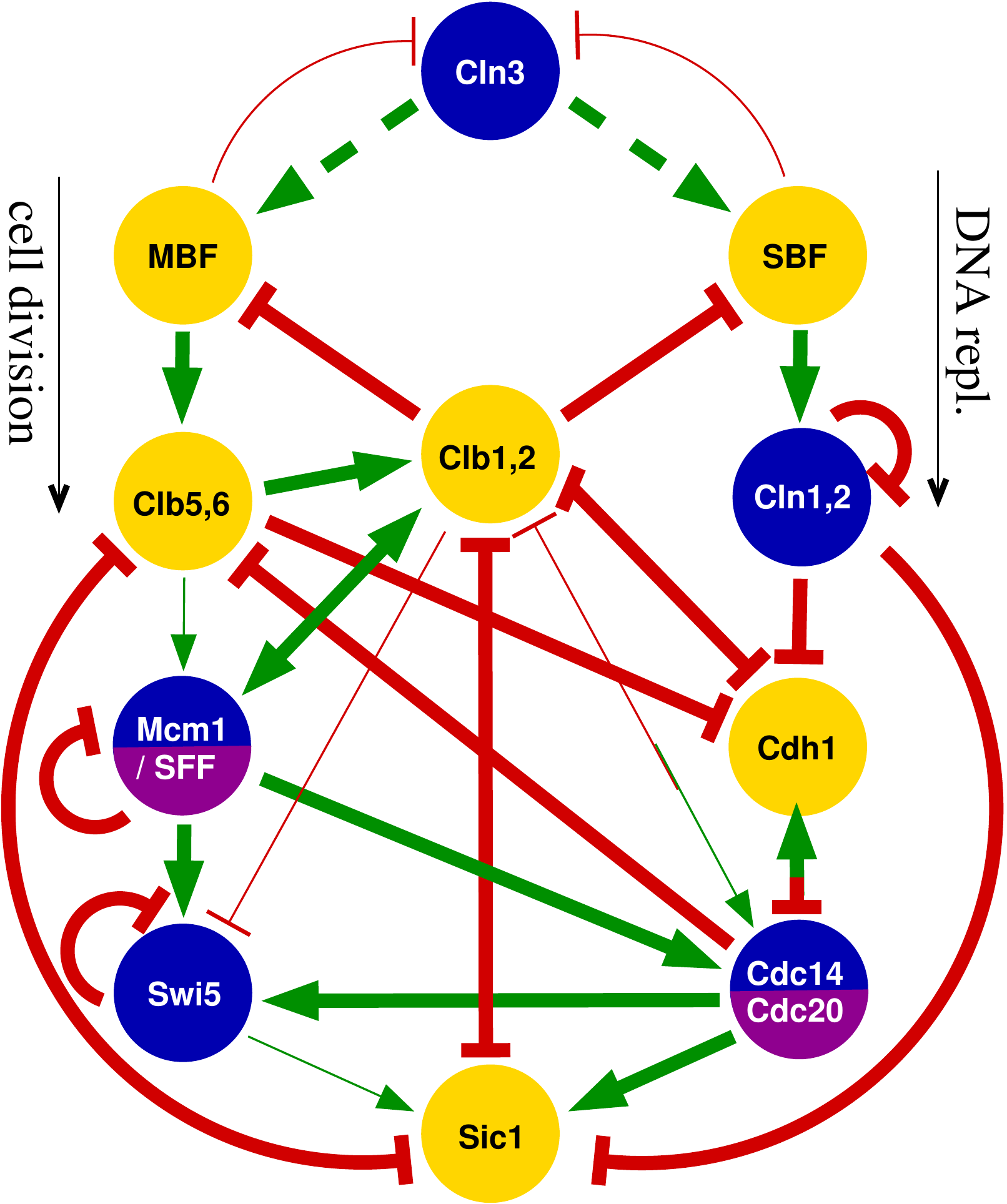}\\
	&\\
	(a) & (b)\\
	\hline
\end{tabular}
\caption{Two model systems for yeast cell cycle. (a) Model $M1$ by Li
  {\em et al.} (b) Model $M2$ by Hong {\em et al.} The weights for
  interactions in $M2$ are 1, 1/3 and 3 for normal, thin and dashed
  lines, respectively.}
\label{models}
\end{center}
\end{figure}

Nevertheless, the small subnetwork responsible for the cell-cycle in
yeast (YCC) is very well studied~\cite{li2004yeast,lee2009robustness,
  okabe2007stable,mangla2010timing, hong2012checkpoints}. A handful of
regulatory proteins or complexes that drive the cell division and the
interactions between them have been identified. Furthermore, several
models of regulatory dynamics involving these proteins have been shown
to yield the experimentally observed expression
profiles~\cite{li2004yeast, mangla2010timing, hong2012checkpoints}. We
will consider two such models here, and demonstrate their high degree
of coherence as a proof of concept.

Once the cell reaches a critical size~\cite{johnston1977coordination,
  rupevs2002checking}, the cell division is initiated. It starts from
the $G_1$ phase of the mother cell and ends in the $G_1$ phases of the
mother and daughter cells. The two models we consider here reproduce
in proper order all the intermediate expression stages that the cell
traverses during division, via the time-evolution equations given in
Table~\ref{timeevolutionequ}. They feature the same set of key
TFs/complexes (vertices of the network) that drive DNA replication and
the accompanying cell division in yeast. These models are depicted in
Fig.~\ref{models}, where blue nodes indicate auto-repressor elements
($c_{ii}=-1$) and two-colored (blue+magenta) nodes represent
multi-protein complexes. The model proposed by Li {\em et
  al.}~\cite{li2004yeast} in Fig.~\ref{models}a is, to our knowledge,
the first successful implementation of Boolean dynamics to yeast's
cell cycle and has been widely used in later studies. We will refer to
this model as $M1$ from now on. More recently, Mangla {\em et al.}
improved $M1$ by introducing additional interactions and varying
interaction weights, while at the same time extending the Boolean
formalism to allow some of the vertices to have three different states
(0,1,2)~\cite{mangla2010timing}. This work was later refined by Hong
et al.~\cite{hong2012checkpoints} in order to capture crucial
checkpoint conditions overlooked in $M1$. We will refer to the YCC
model of Hong {\it et al.} (shown in Fig.~\ref{models}b) as $M2$ from
now on.

Time-evolution equations for $M1$ and $M2$ are given in
Table~\ref{timeevolutionequ}.  Here, $\sigma_i(t)$ is the expression
state which takes the value '$1$' when the node $i$ is active and
'$0$' otherwise, except for Swi5 and Clb2 which can also take the
value of '$2$' (strong expression) in $M2$. The interactions considered in $M1$
(Fig.~\ref{models}a) differ only by their sign, i.e., $c_{ij}\in\{-1,0,1\}$,
while for $M2$ $c_{ij}\in\{0,\pm 1/3,\pm 1, \pm 3\}$ (shown with
different edge thicknesses in Fig.~\ref{models}b). $\theta_{i,1}$ and
$\theta_{i,2}$ are threshold values that trigger the state shifts
$0\leftrightarrow 1$ and $1\leftrightarrow 2$, respectively. While
reproducing the time-evolution of $M2$ given in Ref.\cite{hong2012checkpoints}, we
concluded that the observed dynamics and basin sizes are reproducible
only if an additional parameter, $\delta$, that introduces a
hysteresis to up/down regulation events is added to the evolution
dynamics (see Table~\ref{timeevolutionequ}). This crucial detail
appears to be accidentally omitted in the reference article. We give
the numerical values of the parameters above in
Table~\ref{timeevolutionequ} and refer the reader to the original
references for further details of the two models.

\renewcommand\arraystretch{1.2}
\begin{table}[h!]
	\begin{center}
		\begin{tabular}{|c|c|c|}
			\hline
			\raisebox{-4mm}{$\sigma_i(t+dt)$}&\multicolumn{2}{c|}{if}\\
			\cline{2-3}
			\raisebox{-4mm}{=}&$M1$&$M2$\\
			&with $\theta_{i,1}=0$&with $\theta_{i,1}=0.5$ and $\theta_{i,2}=1.5$\\
			\hline	
			$\sigma_i(t)-1$&$\begin{array}{ll}
			& \sum_{j\neq i} c_{ij} \sigma_j(t) < \theta_{i,\sigma_i(t)}\\
			\mbox{or}&\\
			&\sum_{j\neq i} c_{ij} \sigma_j(t) = \theta_{i,\sigma_i(t)}\ \mbox{and}\ c_{ii}=-1 
			\end{array}$&$\sum_j c_{ij} \sigma_j(t) <\theta_{i,\sigma_i(t)}-\delta$\\
			\hline
			$\sigma_i(t)+1$&$\sum_{j\neq i} c_{ij} \sigma_j(t) >\theta_{i,\sigma_i(t)+1}$&$\sum_{j} c_{ij} \sigma_j(t) \geq \theta_{i,\sigma_i(t)+1}$\\
			\hline
			$\sigma_i(t)$&otherwise&otherwise\\
			\hline
	
		\end{tabular}
	\end{center}
	\caption{Time evolution equations for the two models ($M1$ and
          $M2$) discussed in the text. $\delta=0.75$ (a parameter that
          appears to be omitted in the original reference) was found
          to be consistent with the basin sizes in
          \cite{hong2012checkpoints}.}
	\label{timeevolutionequ}
\end{table}
\renewcommand\arraystretch{1}


While a deterministic implementation of the above equations follows
the experimentally resolved expression dynamics, a stochastic version
will occasionally end up in different attractors listed in
Table~\ref{fixedstates}. The biological relevance of such relatively
low probability states is unclear~\cite{albert2003topology,
  li2004yeast}. Nevertheless, the degree of coherence supported by the
{\em structure} of the regulatory network under the dynamical
evolution given in Table~\ref{timeevolutionequ} should to be defined
through all of its attractors. This also facilitates a fair comparison
of the YCC network with randomized ensembles for which a
classification of attractors according to their biological relevance
is meaningless. On the other hand, we do present numerical results
obtained through basin-size weighted averaging, which, in the case of
YCC, corresponds to essentially focusing on the $G_1$ attractor. Note
also that, the trivial fixed point $\boldsymbol{\sigma}=0$, which is
common to all networks considered in this study, is left out in
Table~\ref{fixedstates} as well as in our analysis below.

\begin{table*}
	
	\begin{center}
		\begin{tabular}{|c|c|cc|}
			
			\hline
			&&&\\
			F&{\includegraphics{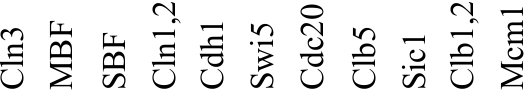}}&\multicolumn{2}{c|}{$\alpha_c$} \\
			&&&\\
			\hline
			&Point attractor in $M1$ \& $M2$&$M1$&$M2$\\
			\cline{2-4}
			&&&\\
			F1& {\includegraphics{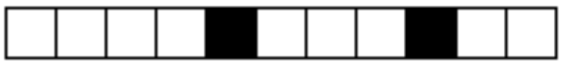}}&1&1\\
			F2 &{\includegraphics{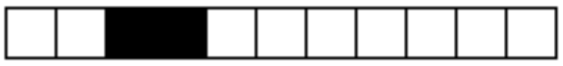}}&1&0.91\\
			F3&{\includegraphics{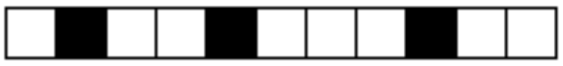}}&0.91&0.91\\
			F4&{\includegraphics{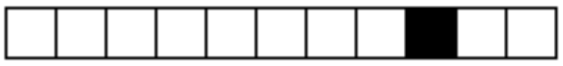}}&1&1\\
			F5&{\includegraphics{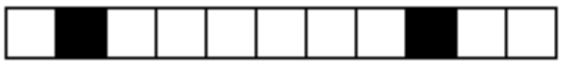}}&0.91&0.91\\
			F6&{\includegraphics{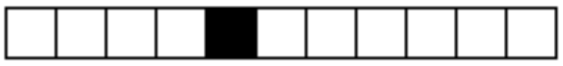}}&1&1\\
			\cline{2-4}
			&Point attractor only in $M2$&&\\
			\cline{2-4}
			F7 &{\includegraphics{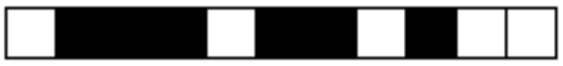}}&-&0.64\\
			F8&{\includegraphics{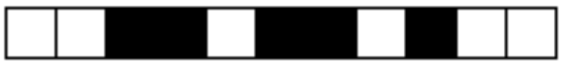}}&-&0.73\\
			\hline
		\end{tabular}
	\end{center}
	
	\caption{YCC attractors under the dynamics given in
          Table~\ref{timeevolutionequ} and their coherence,
          $\alpha_c$. Black boxes indicate the active genes and the
          first row corresponds to the experimentally verified $G_1$
          phase.}
	\label{fixedstates}
\end{table*}

Also given in Table~\ref{fixedstates} are the coefficients of
coherence for the attractors of $M1$ and $M2$, calculated using equation
in Table~\ref{timeevolutionequ}. For the whole network, we obtain
$\alpha_{c} = 0.97$ ($M1$) and $0.89$ ($M2$) with uniformly weighted
attractors and $0.99$ ($M1$) and $0.98$ ($M2$) with
basin-size weighted averaging. We observe that, all the common
attractors of $M1$ and $M2$ (including the biologically relevant $G_1$
phase) are either fully coherent or almost so.

Is the YCC network optimized towards maximal coherence? A meaningful
assessment of the above numbers is possible only in the background of
corresponding results on properly selected random ensembles (for a
similar analysis to determine frequent network motifs in biological
networks see, for example,
Refs.~\cite{alon2006introduction,milo2002network}.) Therefore we next
perform an analysis on directed graphs of the same size with similar
characteristics. We do this in two different ways, as explained below.

\subsection{Coherence: yeast {\it vs} random networks}

Even a network composed of random associations, each gene up- or
down-regulating an arbitrary subset of genes, will display a nonzero
degree of coherence determined by the laws of statistics.  Two obvious
factors that contribute to $\alpha_c$ are: the number of regulating
partners per node, $k$, and the fraction of up-regulating interactions
in the network, $p$. (See section~\ref{sec:alpha_vs_kp}.)
While we will investigate the dependence of
$\bar{\alpha}_{c}$ (avarage of $\alpha_c$ values over an
  ensemble of networks) on $k$ and $p$ below, it is sensible to
compare yeast's cell-cycle network against the appropriate random
ensemble with $(k,p)=(k_{ycc},p_{ycc})$. Excluding self loops, model
$M1$ above has 14 inhibitory and 15 activatory interactions, that is,
$(k_{ycc},p_{ycc})_1 ~\simeq (2.63,0.52)$. Similarly, model $M2$ has 20
inhibitory and 15 activatory interactions, yielding
$(k_{ycc},p_{ycc})_2 ~\simeq (3.18,0.43)$.

\begin{figure}
\begin{center}
\includegraphics[width=\textwidth]{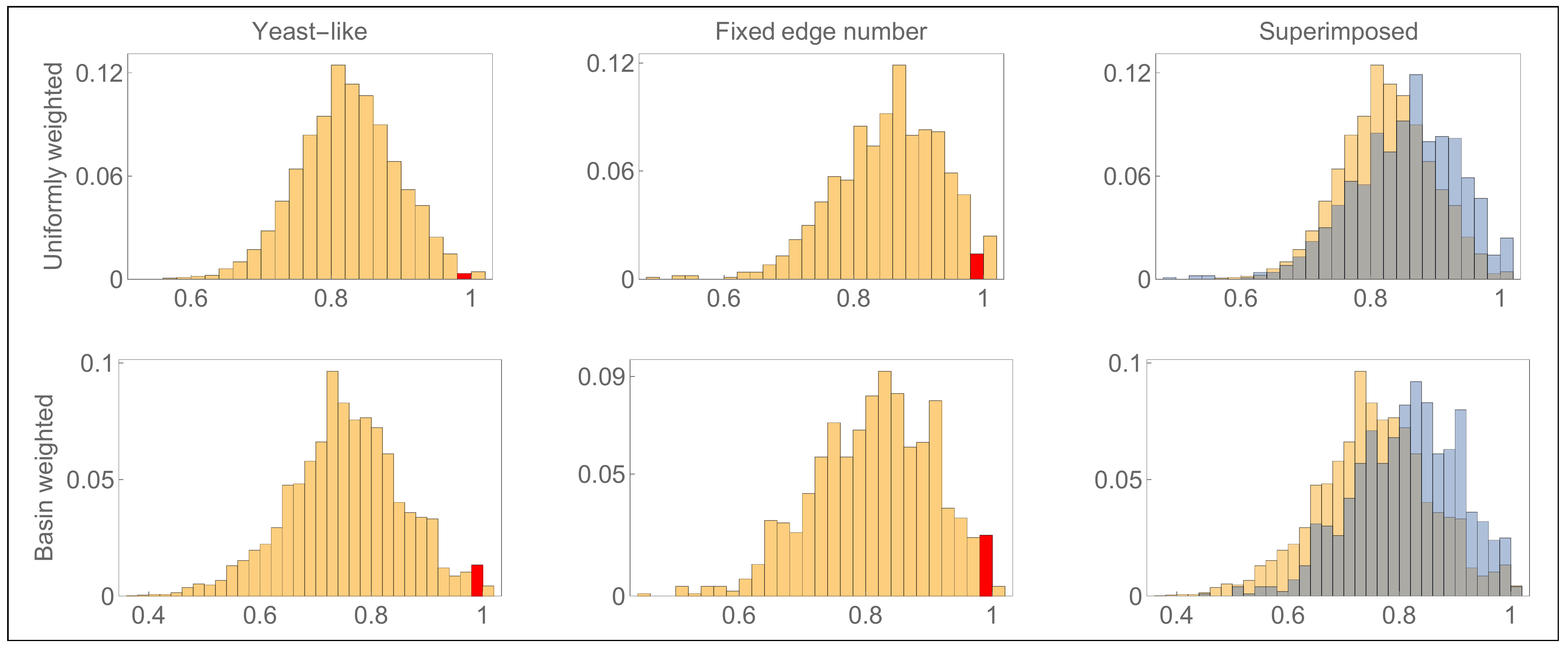}
\caption{The histograms for $\alpha_{c}$ calculated for ``YCC-like''
  networks (first column) and the random $(N_{ycc},k_{ycc},p_{ycc})$
  ensemble (second column) generated by using model $M1$ described in
  the text. The third column shows the superposition of the two
  histograms for comparison. $\alpha_{c}$ for the yeast's cell-cycle
  model is shown in red everywhere.}
\label{fig:alpha_distr}
\end{center}
\end{figure}

\begin{figure}
	\begin{center}
		\includegraphics[width=\textwidth]{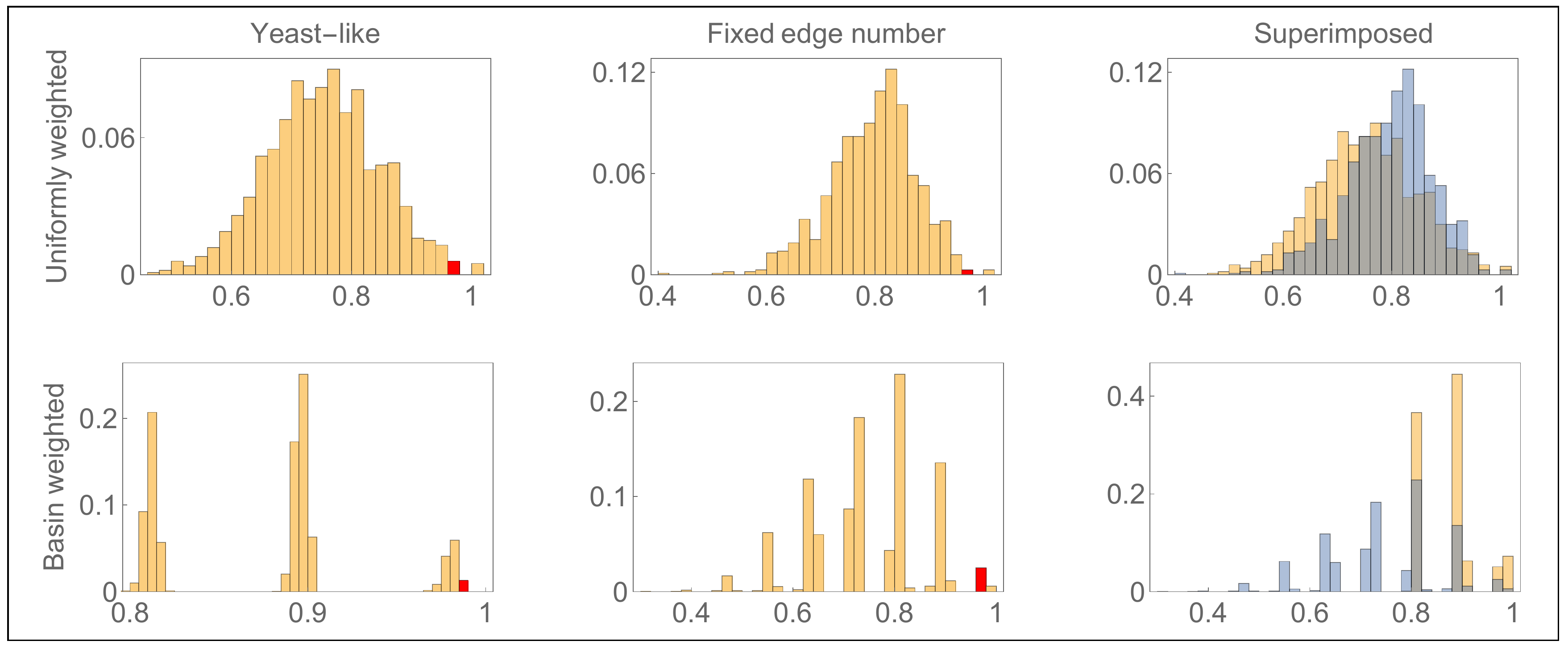}
		\caption{The histograms for $\alpha_{c}$ calculated for ``YCC-like''
  networks (first column) and the random $(N_{ycc},k_{ycc},p_{ycc})$
  ensemble (second column) generated by using model $M2$ described in
  the text. The third column shows the superposition of the two
  histograms for comparison. $\alpha_{c}$ for the yeast's cell-cycle
  model is shown in red everywhere.}
		\label{fig:alpha_distr_Hong}
	\end{center}
\end{figure}

In order to construct an ensemble of networks with fixed $(N,k,p)$, we
first generated connected, undirected random graphs with $N$
regulatory elements (nodes), and $Nk$ interactions (edges) between
them ($k\ge 1$ is required for connectivity.)  Then, each edge was
assigned one of the two possible interaction types (activator or
repressor), such that a fraction $p$ of them up-regulate, and the rest
down-regulate their targets. Note that, a node may act as an activator
for one target and as a repressor for another. This is in line with
the behavior observed in YCC, as well as recent research which
suggests that the TFs involved in both kinds of regulation may be more
common than previously thought~\cite{hickman2007heme,
  ptashne2011principles}. We compared models $M1$ and $M2$ with the
corresponding random ensembles R1 and R2 generated using network
parameters $(N_{ycc},k_{ycc},p_{ycc})_{1,2}$, respectively.

We also applied a more stringent test of coherence in the chosen YCC
models by comparing them with more ``yeast-like'' ensembles. These
ensembles (Y1 and Y2) were generated by forcing each node to preserve
its number of incoming and outgoing edges, {\em separately for each
  type} of interaction. This was achieved by a simple edge-shuffling
procedure that switches the targets of two randomly selected edges in
the network, subject to the condition that the edges are of the same
type.  In all the random ensembles considered, we also separately
fixed the fraction of the auto-repressor nodes ($c_{ii}=-1$) to the
corresponding values 5/11 ($M1$) and 3/11 ($M2$), for a fair comparison.

Overall, two sets of $10^4$ random networks $(R1,Y1)$ and $(R2,Y2)$
were generated for each YCC model. The probability distributions of
$\alpha_{c}$ obtained from each ensemble are shown in
Fig.~\ref{fig:alpha_distr} and Fig.~\ref{fig:alpha_distr_Hong}
separately for $M1$ and $M2$, together with the value obtained from
the actual models for comparison. 

Next, we compared the $\alpha_c$ distibutions above with those from
the ``yeast-like'' ensembles Y1 and Y2, which preserve the particular
set of in-/out-degree pairs for the two interaction types at each node
of the model networks $M1$ and $M2$. Surprisingly, Y1 (both with and
without basin-size weighted averaging) and Y2 (without basin-size
weighted averaging) yield a reduced degree of coherence when compared
with the respective ensembles R1 and R2, as is evident from
Fig.~\ref{fig:alpha_distr}. The corresponding mean values along with standard deviations for  $\bar{\alpha}_{c}$ are listed in Table \ref{meanstd}. We found that the coherence values of the actual networks lie in the top $4\%$ in all of the random ensembles R1, R2, Y1 and Y2. Based on these observations, we conclude that the exceptionally high level of coherent
regulatory activity observed in the two YCC models is not solely due
to by the particular local decoration of the networks' nodes.

\begin{table}[h]
	\begin{center}
		\begin{tabular}{|c|l|c|c|c|c|c|c|}
		\hline
		\multicolumn{2}{|c|}{}&R1&Y1&M1&R2&Y2&M2\\
		\hline
		\multirow{2}{*}{$\bar{\alpha}_c \pm \sigma$ }&{\tiny Uniformly weighted}&$0.85 \pm 0.08$&$0.82 \pm 0.07$&$0.97$&$0.79 \pm 0.07$&$0.75 \pm 0.09$&$0.89$\\
		\cline{2-8}
		&\tiny{Basin weighted}&$0.81 \pm 0.09$&$0.75 \pm 0.10$&$0.99$&$0.75 \pm 0.11$&$0.87 \pm 0.06$&$0.98$\\
		\hline
		\end{tabular}
	\end{center}
	\caption{Mean and standart deviation values for $\alpha_c$ of random ensembles along with the $\alpha_c$ values of model networks $M1$ and $M2$. } 
	\label{meanstd}
\end{table}

\subsection{Coherence {\it vs} network parameters}
\label{sec:alpha_vs_kp}

\begin{figure}[t!]
\begin{center}
\includegraphics[width=\textwidth]{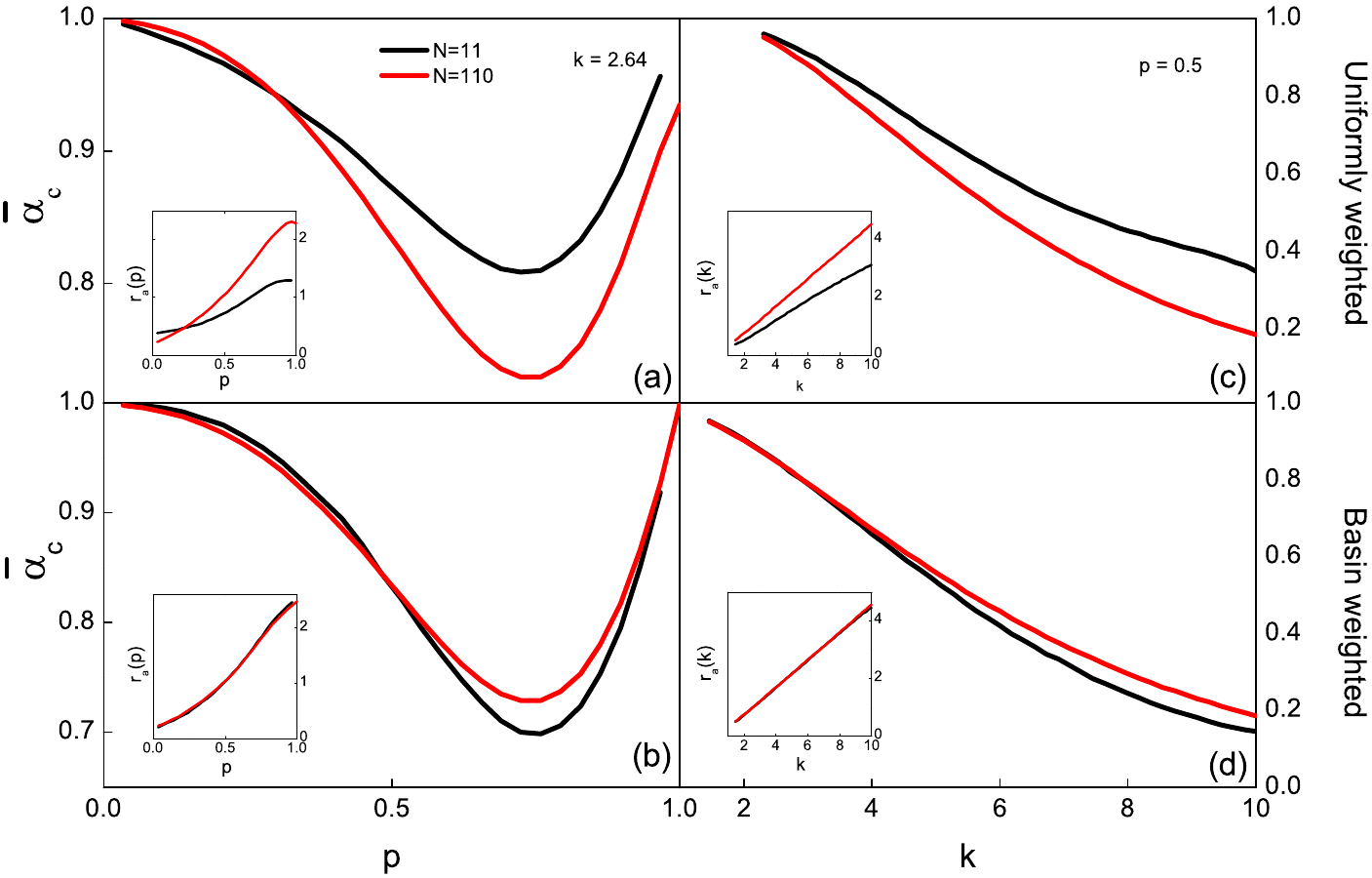}
\caption{(a,b) Degree of coherence, $\bar{\alpha}_{c}$ {\em vs.} $p$
  for $k=k_{ycc}=29/11$ with and without consideration of the basin
  sizes. (c,d) Degree of coherence, $\bar{\alpha}_{c}$ {\em vs.} $k$
  for $p=0.5 \simeq p_{ycc}$ with and without consideration of the
  basin sizes, respectively. The insets show the average fraction of
  the regulating active nodes at the fixed points.}
\label{fig:alphabar}
\end{center}
\end{figure}

In order to investigate the dependence of mean degree of coherence on
$k$ and $p$, we next considered (for simplicity of implementation)
only the YCC model $M1$ above. We, again, generated $10^4$ distinct,
random networks of size $N=11$ (as in YCC), now with $k \in [1.6,5]$ and 
$p=0.5\simeq p_{ycc}$, as well as with $k=k_{ycc}=29/11$ and $p \in
[0,1]$. Possible isomorphism of the ensemble members was ruled out by
ensuring that each network in the ensemble has a different eigenvalue
spectrum. We also generated much larger networks with $N=110$ nodes
(corresponding to a 10-fold increase in size) for comparison and with
$k$ and $p$ values in the same intervals as above, in order to check
the network size dependence of our results. Then the attractors of the
dynamics given in Table~\ref{timeevolutionequ} were found for each
network by an exhaustive search for $N=11$ and by sampling $2^{11}$
initial states for $N=110$. Finally, $\alpha_{c}$ were calculated
separately for each network over corresponding attractors of the
dynamical evolution described by Eq. (\ref{eq:coherent}). The dependence of
the ensemble mean $\bar{\alpha}_{c}$ on $p$ and $k$ is shown in
Figs.~\ref{fig:alphabar}(a)\&\ref{fig:alphabar}(b) respectively.

$\bar{\alpha}_{c}$ changes only slightly with the number of nodes $N$
for a given $(k,p)$ pair: increasing the network size 10-fold yields a
similar behavior, with $<9.8 \%$ difference in the worst case
(Fig.~\ref{fig:alphabar}). Therefore, despite the smallness of the
model system considered here, our results on random ensembles may be
expected to serve as a reasonable null-hypothesis for coherence in the
global regulatory network of yeast and other organisms.

It may at first sight be surprising that $\bar{\alpha}_c$ is
asymmetric with respect to $p=1/2$, given that the number of genes
with incoming edges that both up- and down-regulate the gene is
maximized precisely at this point. The observed asymmetry is due to
the fact that the fraction of active nodes at the fixed points is a
monotonically increasing function of $p$ (the insets of
Fig.~\ref{fig:alphabar}). Close to $p=0$ and $p=1$, most genes are
regulated in the same direction by their regulating partners, hence an
increase in coherence is observed in these extremes. While the
fraction of genes with conflicting incoming edges peaks at $p=1/2$,
the minimum of $\bar{\alpha}_{c}$ is shifted towards larger $p$ values
(more up-regulation), where the chances are higher for antagonistic
regulatory partners of a gene to be simultaneously expressed.

\begin{figure}
	\begin{center}
		\includegraphics[width=10cm]{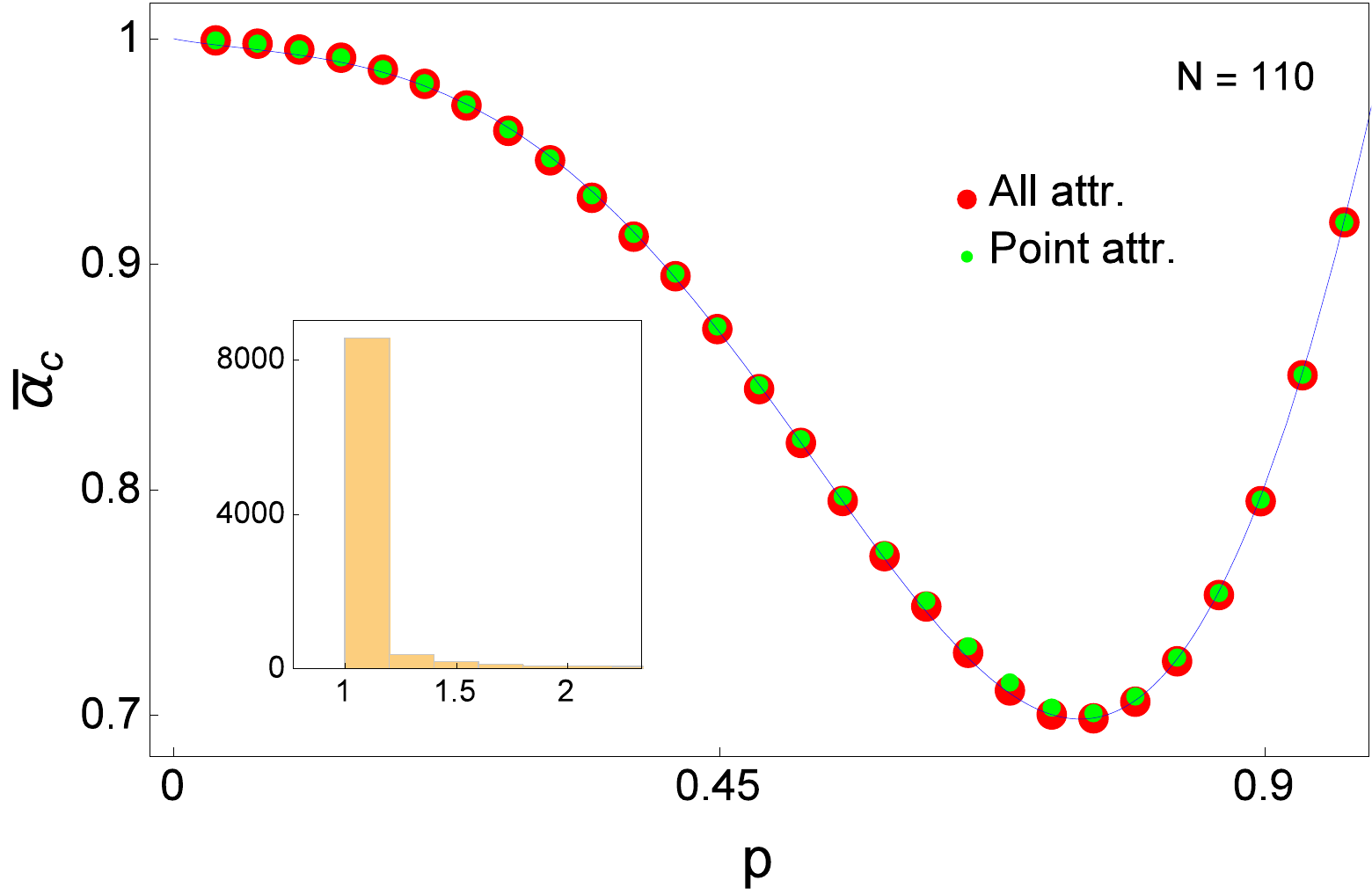}
	\end{center}
	\caption{The contribution of cyclic attractors for $N=110$ to
          basin-size weighted values of $\bar{\alpha}_c$. The
          dependence of $\bar{\alpha}_c$ on $p$ is calculated over all
          attractors (red) and over point attractors only (green), for
          comparison. The blue curve is a fit calculated by using
          equations (\ref{eq:r_a}\&\ref{eq:a_c}). The average cycle
          length distribution for $p=0.5$ given in the inset is
          typical of all the ensembles considered here, irrespective
          of the averaging method used.}
	\label{alpha_p_cycle}
\end{figure}

A simple analytical model based on the above reasoning can be shown to
reproduce the observed behavior. Having shown the weak dependence of
our results on the network size, let us consider the large-$N$
limit. For fixed $(k,p)$, define the average number of {\em active}
regulating partners per node be $r_a(k,p)$. The data from the
ensembles with $N=11$ and $N=110$ suggest the form below (see
Fig.~\ref{fig:alphabar}(a,b) inset):
\begin{equation}
\label{eq:r_a}
r_a(k,p) = r_{min}+p^\gamma(r_{max}-r_{min})
\end{equation}
where $r_{min,max}$ are $k$-dependent.  Note that, a typical attractor
of a network with all inhibitory interactions ($p=0$) will still have
a few nodes ``on'' in absence of active down-regulating partners,
hence $r_{min}>0$. When $p=1$, we expect most of the nodes to be
turned on. A node is coherently regulated, if all its {\it active}
regulating partners act unanimously. $\alpha_{c}$ measures essentially
the probability for this event across the network's
attractors. Treating the states of neighbors regulating a given node
as independent random variables, this probability can be expressed as
\begin{eqnarray}
\label{eq:a_c}
\bar{\alpha}_{c}(p,k) = p^{r_a} + (1-p)^{r_a}\ .
\end{eqnarray}
The functional form in equation (\ref{eq:r_a}) yields a good fit for
both $r_a$ and $\bar{\alpha}_{c}$ with $(r_{min},r_{max},\gamma)$
values listed in Table \ref{datafit}. The agreement between the
basin-averaged $\bar{\alpha}_c$ {\em vs} $p$ for $N=110$ and the
analytical form above is shown in Fig.~\ref{alpha_p_cycle}.
On the other hand, the parameter values yielding the best fit to
$\bar{\alpha}_c$ and $r_a$ differ. This discrepancy may be attributed
to the nonuniformity of the in-degree distribution and the
contribution of correlations, both of which are not captured by the
simple treatment above.

\begin{table}
	\begin{center}
\begin{tabular}{cc|c|c|c|c|c|c|}
	\cline{3-8}
	& & \multicolumn{3}{ |c| }{for $r_{a}$}& \multicolumn{3}{ c| }{for $\bar{\alpha}_c$} \\ \cline{3-8}
	& & $r_{min}$& $r_{max}$ & $\gamma$ & $r_{min}$ & $r_{max}$ & $\gamma$ \\ \cline{1-8}
	\multicolumn{1}{ |c  }{\multirow{2}{*}{N=11} } &
	\multicolumn{1}{ |c| }{no basin} & 0.36 & 1.46 & 1.49 & 1.03 & 1.66 & 1.91\\ \cline{2-8}
	\multicolumn{1}{ |c  }{}                        &
	\multicolumn{1}{ |c| }{with basin} & 0.19 & 2.69 & 1.54 & 1.02 &2.55 & 2.64     \\ \cline{1-8}
	\multicolumn{1}{ |c  }{\multirow{2}{*}{N=110} } &
	\multicolumn{1}{ |c| }{no basin} & 0.21 & 2.55 & 1.44 & 1.03 & 2.26 & 2.42 \\ \cline{2-8}
	\multicolumn{1}{ |c  }{}                        &
	\multicolumn{1}{ |c| }{with basin} & 0.22 & 2.60 & 1.49 & 1.03 & 2.23 & 2.34 \\ \cline{1-8}
\end{tabular}
\end{center}
\caption{Fit values for the model.}
\label{datafit}
\end{table}

The monotonic decrease of $\bar{\alpha}_c$ with increasing $k$, the
mean number of regulating partners per node, shown in
Fig.~\ref{fig:alphabar}(c,d) is hardly surprising. It reflects the
fact that the genes with fewer active regulators are less likely to
receive conflicting inputs. The number of active neighbors per node
when $p=0.5\simeq p_{ycc}$ increases almost linearly with $k$,
suggesting that the overall number of simultaneously active nodes
remains a constant, essentially determined by $p$.

The results presented here are identical to the naked eye when
non-point-attractors (with period $>$ 1) are excluded from the
calculation of $\alpha_{c}$ on random networks. Such cycles appear
relatively infrequently with the given regulatory dynamics (see
Fig.~\ref{alpha_p_cycle} inset). Therefore, the fact that the yeast's networks
in (Fig.~\ref{model}) has only point attractors does not seem to be
relevant to its extreme degree of coherence, either.
Investigation of structural features of coherent networks (such as local
motif statistics, directed cycles, etc.) is planned as a future study.

\section{Discussion}
\label{discussion}
We defined a measure for the degree of coherence in gene regulatory
networks. Using the proposed measure, we showed that the cell-cycle
network of the budding yeast displays exceptionally high degree of
harmonious regulatory activity, which is in line with earlier
observations on its robustness. The proposed measure of coherence sets
the YCC network aside, within an ensemble of networks of the same size
and composition. We found that, achieving coherent regulation is most
difficult in systems when roughly 25\% of the interactions are
repressory. While this ratio is about 1/2 for the yeast's cell-cycle
network, YCC yields a degree of coherence which is significantly
higher than those of random networks with identical network parameters
such as size, average degree, and +/- interaction ratio. When basin
basin sizes are ignored (uniform weighting), optimality of the YCC
network is even more pronounced within the ensemble of yeast-like
networks that preserve the in-/out-degrees of nodes separately for
each interaction type. A deeper analysis of highly coherent random
networks is necessary to pinpoint the structural determinants of these
unique and possibly biologically relevant class of networks.

Perhaps the most far-reaching question inspired by our findings is
whether coherence in gene regulation is a prevalent motif across
organisms in nature. In order to check this, the presented analysis
needs to be extended to the cell-cycle regulation networks of other
organisms (for example the fission yeast \cite{davidich2008boolean}),
as well as GRNs responsible for different biological processes
(examples in the literature include cell differentiation and
segmentation
networks~\cite{remy2008minimal,mendoza1999genetic,krumsiek2011hierarchical}). Such
an investigation is currently in progress and will shed light on the
generality of our observations on the cell-cycle network of the
budding yeast.
An affirmative answer to the question above would point to the
interesting possibility of a Hebbian-like selection mechanism for
regulatory interactions. That is because, the coherence property
defined in this work can be seen as a generalization of the Hebb's
rule for neural networks~\cite{}, which is usually stated as ``neurons
that fire together, wire together''. A mechanism of selection for
coherent networks would similarly promote positive regulation (and
demote negative regulation) among genes that are expressed
simultaneously, and the vice versa for an ``on'' gene regulating an
``off'' gene. Rapid evolution of regulatory interactions as found in
recent studies~\cite{schmidt2010five} resonates with this scenario,
for it provides the fertile ground on which such a mechanism can
effectively produce coherent networks in short evolutionary time
scales.

\section*{Acknowledgements}
A.K. is grateful for the hospitality of M. Kardar during a visit to
MIT and a stimulating discussion with the rest of the biophysics group
there. We also thank D. Yuret and C. Dunn of Ko\c c University for
beneficial exchanges.

\section*{References}
\bibliography{manuscript}

\appendix

\end{document}